\documentstyle[prb,aps,psfig,floats]{revtex}
\newcommand{\beq}{\begin{equation}}
\newcommand{\eeq}{\end{equation}}

\newcommand{\bfi}{\begin{figure}}
\newcommand{\efi}{\end{figure}}

\begin{document}

\title{{\bf Interplay between superconductivity 
and flux phase in the $t$-$J$ model}}  
\vspace{2cm}
\author{E. Cappelluti and R. Zeyher} 
\vspace{1cm}

\address{Max-Planck-Institut\  f\"ur\
Festk\"orperforschung,\\ Heisenbergstr.1, 70569 Stuttgart, Germany}

\vspace{3cm}

\maketitle

\begin{abstract}
We study the phase diagram of the $t$-$J$ model using a mean field type
approximation within the Baym-Kadanoff perturbation expansion for
Hubbard $X$-operators. The line separating the normal state from
a $d$-wave flux or bond-order state starts near optimal doping at $T=0$
and rises quickly with decreasing doping. The transition temperature $T_c$ 
for $d$-wave superconductivity increases monotonically in the overdoped 
region towards optimal doping. Near optimal
doping a strong competition between the two $d$-wave order parameters 
sets in leading to a strong suppression of $T_c$ in the underdoped region. 
Treating for simplicity the flux phase as commensurate the superconducting 
and flux phases coexist in the underdoped region below $T_c$, whereas 
a pure flux phase exists above $T_c$ with a pseudo-gap of $d$-wave 
symmetry in the excitation spectrum. We also find that incommensurate 
charge-density-wave ground states due to Coulomb interactions do not 
modify strongly the above phase diagram near the superconducting phase, 
at least, as long as the latter exists at all.

\par
PACS numbers: 74.20.Mn, 74.25.Dw., 71.27.+a
\end{abstract}

\section{Introduction}
\label{introsec}

It is widely accepted that the mechanism causing
high-$T_c$ superconductivity in the cuprates is intimately
related to the underlying properties of the normal state.
However, there exists presently no agreement on what 
an appropriate and correct description of the normal state is. 
One reason for this is that experiments indicate the
presence of a whole variety of important fluctuations 
in the normal state. At low dopings antiferromagnetic fluctuations
dominate \cite{rossat-mignod,pines}. The associated 
quantum critical point, however, lies far apart from optimal doping 
so that the relevance of magnetic fluctuations for high-$T_c$ 
superconductivity is not evident.
Other fluctuations which have been considered to be relevant in the
normal state are associated with 
resonance-valence bonds \cite{anderson1,affleck1,fukuyama,ubbens}, flux phases
\cite{affleck2,wen,lee1}, stripes \cite{emery1,tranquada}, or charge density 
waves \cite{castellani1,markiewicz}.
The close proximity of high-$T_c$ superconductivity and structural 
phases or cross-overs also follows from the experimental observation
of a spin-gap or pseudo-gap of $d$-wave symmetry in the underdoped regime 
of the cuprates \cite{loram,cooper,ding,loeser,boebinger}. 
It is thus desirable to study in more detail the
phase diagram of models relevant to the cuprates and to identify
instabilities, their symmetries and their interactions. \par

The low energy physics of the $Cu O_{2}$ layers
of the cuprates is well described by the $t$-$J$ model 
\cite{anderson1,zhang1}.
In order to be able to carry out systematic approximations
the two spin degrees of freedom per site are often increased to
$N$ degrees of freedom so that $1/N$ can be used as a small parameter 
\cite{kotliar1,ruckenstein}. 
For our purposes it is convenient to consider two spin degrees of 
freedom as in the original model and to extend the number of orbitals
per site from one to $N/2$. In this way the symmetry group of the
Hamiltonian of the $t-J$ model is enlarged from $SU(2)$ to
the symplectic group $Sp(N/2)$. Based on $1/N$ expansions 
it has been shown that this model possesses at large $N$'s two intrinsic 
instabilities\cite{wang}.
The first one \cite{grilli1,greco,zeyher1} is associated with $d$-wave 
superconductivity and is
obtained in $O(1/N)$. The second one \cite{zeyher1,morse,grilli2,ercolessi}
is related to a flux or bond-order wave
state of $d$-wave symmetry and is obtained already in the leading order
$O(1)$. The instability towards superconductivity is found for all
dopings; the corresponding transition temperature $T_c$ increases
from the overdoped side with decreasing doping. Near and below optimal doping
the second instability sets in. It occurs in the $d$-wave channel for
the relevant region of parameters. Both its location and its symmetry
fits well to the experimental observations of a $d$-wave pseudogap
in the underdoped region. It is thus of high interest to study
these two intrinsic instabilities of the $t-J$ model 
in more detail and this is the subject of this paper. \par

One obstacle for such a study is the fact that the two instabilities
occur in different orders of the $1/N$ expansion. Whereas the transition
between the normal and charge-density wave or flux states can be 
obtained by comparing $O(1)$ contributions to the free energy the 
superconducting part is of $O(1/N)$. Nevertheless, the calculation
of the superconducting phase boundary is unique and involves only
Green's functions of leading order in the $1/N$ expansion.
The position of the boundary in the phase diagram, on the other hand,
depends on the value of $N$. We will use the $1/N$ expansion in the sense
that it allows to select a set of diagrams which are the leading ones
at least at large $N$'s. After this selection has been made we put $N=2$.
The anomalous self-energy of $O(1/N)$ contains quite a number of terms.
It has recently been shown that only the $d$-wave contributions
are important and that for this symmetry the retarded terms cancel each
other to a large extent. To simplify our treatment we will
thus only keep instantaneous terms in the $d$-wave channel for the
superconducting part. \par
In the following we will enforce the constraints of the $t-J$ model
by using $X$-operators\cite{ruckenstein,zeyher2,zeyher3}. Such a formulation is 
not more involved than the
more familiar one using a slave boson formulation. It has, however, the
advantage that all quantities such as the order parameters
are gauge invariant in the sense of slave boson theory. 
The outline of the paper is the following: in 
section \ref{meansec} the formalism of our approach 
is introduced, and in section \ref{incomsec} we discuss
the instability of the normal state towards incommensurate
and commensurate flux phases, clarifying the range of validity
of treatments which only take commensurate flux phases into account.
The interplay
between superconductivity and flux phase is discussed 
in detail in section \ref{supersec}. Finally we investigate
in section \ref{pssec} the role of phase 
separation and the effect of 
Coulomb interaction on the phase diagram.

\section{Mean-field equations in terms of $X$-operators}
\label{meansec}

Using $X$-operators the Hamiltonian of our generalized $t-J$ model
has the form

\begin{eqnarray}
H & = & - \sum_{i j \atop p=1 \ldots N}
\frac{t_{i j}}{N} X_{i}^{p 0} X_{j}^{0 p} +
\sum_{i j \atop p,q =1 \ldots N}
\frac{J_{i j}}{4N} X_{i}^{p q} X_{j}^{q p}\nonumber\\
 & & - \sum_{i j \atop p,q =1 \ldots N}
\frac{J_{i j}}{4N} X_{i}^{p p} X_{j}^{q q}+
\sum_{i j \atop p,q =1 \ldots N}
\frac{V_{i j}}{2 N} X_{i}^{p p} X_{j}^{q q}\:.
\label{htj}
\end{eqnarray}
$i$ and $j$ run over the sites of a lattice. For $N=2$ the operators
$X_i^{pq}$ are identical with the projection operators 
$|{p \atop i}><{q \atop i}|$, where $|{p \atop i}>$ denotes for
$p=0$ the empty state and for $p=1,2$ singly occupied states at site $i$
with spin up and down.
$t_{ij}$ and $J_{ij}$ are hopping and exchange constants, respectively,
and both are assumed to act only between nearest neighbors. The last
term in Eq.(1) describes the Coulomb repulsion between electrons.
\par

The extension from $N=2$ degrees of freedom per site to a general
$N$ is accomplished by introducing a flavor index $\mu=1,...,N/2$
which enumerates $N/2$ copies of the original orbital. The index $p$
is then a composite index $p=(\sigma,\mu)$, where $\sigma$ denotes
a spin index, and can be chosen 
to run from $0$ to $N$. The $X$-operators are in general no longer 
projection operators but 
are assumed to obey still the commutation rules
\begin{equation}
\left[ X_{i}^{pq}, X_{j}^{rs}\right]_{\pm} =
\delta_{ij} \left( \delta_{qr} X_{i}^{ps} \pm
\delta_{sp} X_{i}^{rq} \right) \: .
\label{algebra}
\end{equation}
The upper (lower) signs in Eq.(2) hold for fermionlike (bosonlike 
or of mixed nature) $X$-operators. Per definition, fermionlike 
operators have the internal indices $p = 0$, $q > 0$ or $q = 0$, $p > 0$,
bosonlike ones $p = q = 0$ or $p, q > 0$. Moreover, the
diagonal $X$-operators are assumed to obey the constraint
\begin{equation}
X_{i}^{00} + \sum_{p=1\ldots N} X_{i}^{pp} = 
\frac{N}{2} \:.
\label{complete}
\end{equation} 
In the usual case $N=2$ both Eqs.(2) and (3) are fulfilled,
and Eq.(3) is just the completeness relation. For a general $N$ Eq.(3)
means that at most $N/2$ electrons can occupy the $N$ states
at site i since $X_i^{00}$ is a non-negative operator\cite{zeyher3}.
Moreover, it can be shown that the problem is completely specified if
one assumes, in addition, that the diagonal operators $X_i^{pp}$ are projection
operators for $p>0$, exactly as in the case $N=2$\cite{zeyher3}.
Before proceeding we want to  introduce the
notation we will use in the following. The index $1$ in
$X(1)$ denotes all the degrees of freedom 
(internal indices $p_1,q_1$, imaginary time $\tau_1$, site index $i_1$) 
of the $X$-operator, so that
$1 = ( \begin{array}{c}
p_{1}q_{1}\\
\bar{1}
\end{array})$ where
$\bar{1}$ stands for $\bar{1} = (i_{1}, \tau_{1})$.

Following the Baym-Kadanoff formalism\cite{baym,ruckenstein,zeyher2}, we define
the single particle non-equilibrium Green's function in 
the presence of an external source $K$ by 
\begin{equation}
G(12) = - <TSX(1)X(2)>/<S>,
\end{equation}

\begin{equation}
S = T \exp\left[ \int d1 K(1)X(1) \right].
\end{equation}
In Eqs.(4) and (5) $T$ is the time ordering operator and
$K^{pq}$ the external source which is assumed to couple only to bosonlike 
operators $X^{pq}$. $\int d1$ means $\sum_{p_1,q_1,i_1} \int_0^\beta d \tau_1$
where $\beta$ is the inverse temperature. 
It is convenient to introduce a normalized Green's
function $g$
such that Dyson's equation has the usual delta-function
on the right side:\cite{zeyher2,kulic}
\begin{equation}
\int d2 (G^{-1}_0(12) - \Sigma(12))g(21') = \delta(1-1'),
\end{equation}
where $G_0$ is given by
\begin{equation}
G^{-1}_0(12) = \delta(1-2){\partial \over{\partial \tau_2}}
-\delta({\bar 1}-{\bar 2})(K^{00}({\bar 1}) \delta_{q_1q_2}
-K^{q_1q_2}({\bar 1})),
\end{equation}
and
\begin{equation}
\Sigma(12) = -\int d3 v(132) <X(3)> +
\int d3 d4 d5 v(134) g(45) \gamma(52;3)\:.
\label{selg}
\end{equation}
The explicit expression for the function $v$ is 
$$
v(123) = \delta(\tau_2-\tau_1) \delta(\tau_3-\tau_1) \left[
(t_{i_1i_3} +J_{i_1i_2}/2 \delta_{i_1i_3})(\delta_{q_10} \delta_{q_20}
(1-\delta_{p_30})\delta_{p_2p_1} \delta_{q_2p_3} \right.  \nonumber $$ 
$$  - \delta_{p_10}
\delta_{p_30}(1-\delta_{q_30}) \delta_{p_2q_3} \delta_{q_1q_1}) 
+(t_{i_1i_3} +(J_{i_1i_2}/2-V_{i_1i_2})\delta_{i_1i_3})
\nonumber $$
\begin{equation}
(\left.\delta_{q_10} \delta_{q_30}(1-\delta_{p_30})\delta_{p_20} 
\delta_{q_2q_1}
\delta_{p_3p_1} -\delta_{p_10} \delta_{p_30}(1-\delta_{q_30})
\delta_{p_2p_1} \delta_{q_20} \delta_{q_3q_1}) \right].
\label{vfu}
\end{equation}
The square bracket on the right-hand side of Eq.(\ref{vfu}) contains two
contributions: The first one describes the hopping of an hole
between nearest neighbors 
with a spin-flip, the second one without a spin-flip.

The mean field approximation is usually defined as an approximation
for the self-energy $\Sigma$ in which $\Sigma$ contains all skeleton 
diagrams which are at most linear in $g$. This means in our case
that the vertex $\gamma$ should be approximated by the contribution
of zeroth order in $g$\cite{sandalov}. 
Such an approximation, however, would violate Luttinger's theorem for 
every $N$. Moreover,
the relation between the number of particles at a site and the
chemical potential would be not unique and depend on the way one calculates
it. One way to get rid of these artefacts is to keep in the normal
part of $\Sigma$ only those terms which also are present in the large-$N$
limit and then to put $N=2$. Another justification for such 
a procedure is the following: The leading normal and anomalous parts
 of $\Sigma$ in Eq.(\ref{selg}) are of $O(1)$ and $O(1/N)$, respectively, in the 
$1/N$ expansion. The calculation of the superconducting transition 
temperature involves in leading order only normal self-energies of
$O(1)$, which, however, may already describe modulated phases. Calculations
based on $1/N$ expansions thus suggest that first one should solve the $O(1)$
normal state problem in leading order of the $1/N$ expansion which
may already imply the consideration of modulated phases. In a second
step the solution is then used to solve the superconducting part
in leading order in $1/N$. It has been found that the mean-field
expression for the anomalous $\Sigma$ is in general quite a good
approximation for the total $O(1/N)$ contribution to the anomalous 
self-energy. Taking this for granted one arrives exactly at the above
procedure to calculate $\Sigma$. \par
Before working out the details of the above  approximation for $\Sigma$ 
it is useful to summarize the $O(1)$ results for $\Sigma$ and $g$
in the normal, unmodulated phase.
In leading order of the $1/N$ expansion 
the vertex $\gamma$ in the
self-energy can be approximated by the bare one, i.e., by
\begin{equation}
\gamma(12;3)=\delta(\bar{1}-\bar{2})
\delta(\bar{2}-\bar{3})
[ \delta_{p_{1}0}\delta_{p_{2}0}
\delta_{p_{3}q_{1}}\delta_{q_{3}q_{2}}-
\delta_{q_{1}0}\delta_{q_{2}0}
\delta_{q_{3}p_{1}}\delta_{p_{3}p_{2}}]\:.
\label{gamma}
\end{equation}
Also in leading order of the $1/N$ expansion the expectation value of 
bosonic operators can be calculated in the absence of source field via
\begin{equation} 
<X^{pq}({\bar 1})> = \delta_{pq}g({{0q} \atop {\bar 1}}{{q0} 
\atop {{\bar 1}^+}}).
\end{equation}
Inserting Eq.(\ref{gamma}) into Eq.(\ref{selg}) 
the self-energy is frequency-independent
for zero source fields. Denoting it after a Fourier transform by
$\epsilon({\bf k})$ we obtain for $g$ 
\begin{equation}
g({\bf k},i\omega_{n}) = 
\frac{1}{i\omega_{n}-\epsilon({\bf k})-\lambda+\mu}\:,
\end{equation}
with
\begin{equation}
\epsilon({\bf k}) =
\delta \epsilon_{0}({\bf k}) + \alpha({\bf k})\:.
\end{equation}
$\epsilon_{0}({\bf k})$ is the free electron dispersion
$\epsilon_{0}({\bf k}) = -t({\bf k})/2$, and the doping $\delta$,
the function $\alpha$, and the shift $\lambda$ are 
determined by the following expressions:
\begin{equation}
n = 1 - \delta =
\frac{2}{N_{c}} \sum_{{\bf p}} f\left[
\frac{\epsilon({\bf p})}{T}\right]\,,
\end{equation}
\begin{equation}
\lambda = \frac{1}{N_{c}} \sum_{{\bf p}} t({\bf p})
\eta({\bf p})-\frac{1}{2}\left[\frac{J(0)}{2}-
V({\bf q} \rightarrow 0)\right] n
\label{la1}
\:,
\end{equation}
\begin{equation}
\alpha({\bf k}) = - \frac{1}{2 N_{c}} \sum_{{\bf p}} 
J({\bf k+p})
\eta({\bf p})\:.
\end{equation}
$N_{c}$ is the total number of sites, $f(x)$ the Fermi function $f(x) = 
1/(\exp(x)+1)$, 
and $\eta({\bf k}) = f[\epsilon({\bf k})/T]$. $J(\bf q)$ is defined by
$J({\bf q}) = 2J ( \cos(q_x)+\cos(q_y))$, where $J$ is the 
coupling constant $J_{ij}$ for nearest neighbor $i,j$. 
The long-range Coulomb interaction $V_{i j}$,
written in the Fourier space, has the expression\cite{becca}:
\begin{equation}
V({\bf q}) = \frac{V_{C}}{2 \sqrt{A(q_{x},q_{y})^{2}-1}}\:,
\label{v(q)}
\end{equation}
with $A(q_{x},q_{y}) = c[\cos(q_{x})+\cos(q_{y})-2]-1$.  The
constant $c$ has been estimated to be around 50. According to
Eq.(\ref{v(q)}) the 
long-range nature of $V$ implies that
$\lim_{{\bf q} \rightarrow 0} V({\bf q}) = \infty$. In equilibrium
this infinite large constant is compensated by the lattice due to
charge neutrality. For a nonequilibrium value of the density this
compensation no longer works. 
As a consequence, the isothermal compressibility, defined by
$\kappa = n^{2} (\partial \mu /\partial n)$, becomes
\begin{equation}
\kappa = \kappa_0 + \frac{V({\bf q}\rightarrow 0)}{2}\, n^{2} > 0\:,
\label{isokappa}
\end{equation}
where $\kappa_0$ means the compressibility without the Coulomb interaction.
Whatever the value of $\kappa_0$ is, the total $\kappa$ is according
to Eq.(\ref{isokappa}) always positive preventing any phase separation.

\section{Commensurate and incommensurate\newline flux phases}
\label{incomsec}

The expressions for normal state quantities derived above are valid as 
long as the doping is not too small, where 
the normal state is unstable with respect to other phases\cite{grilli1}.
In particular, the instability towards bond-order states and flux phases
has been investigated as  
function of the coupling $J/t$\cite{affleck2,grilli2}, and the 
temperature $T$\cite{ubbens,ercolessi}.
In these works, the new phases have been assumed to be commensurate 
with a commensurate modulation vector ${\bf Q}_{c} = (\pi,\pi)$.
(We use here the terms ``commensurate'' and ``incommensurate''
with respect to the lattice periodicity and not, as in
Refs. \onlinecite{lederer,poilblanc} with respect to the 
electronic filling).
On the other hand, numerical studies\cite{lederer,poilblanc}
as well as a zero temperature slave boson calculation\cite{morse}
 indicated that
an incommensurate flux state can be more
stable than a commensurate one for a general doping. In this section
we study this problem in more detail within our approach, especially, also 
at finite temperatures.

The generalization of Dyson's equation for the Green's 
function
$g({\bf k},i\omega_{n})$ to the non-periodic case is
\begin{equation}
(i\omega_{n} + \mu) g({\bf k},{\bf q},i\omega_{n})-\frac{1}{N_{c}} \sum_{{\bf 
p}}
\Sigma({\bf k,p}) g({\bf k-p},{\bf q-p},i\omega_{n})=
N_{c}\delta({\bf q})\:,
\label{dysonun}
\end{equation}
where we have defined the Fourier transformation by

\begin{equation}
g({\bf k,q},i\omega_n)= \sum_{i,j} g(i,j,i \omega_n)
e^{i{\bf k}\cdot({\bf R}_{i}-{\bf R}_{j}) +i{\bf q}\cdot{\bf R}_{j}}\:.
\end{equation}
${\bf R}_i$ denotes the lattice vector to the site $i$. 
The instability towards an incommensurate state causes 
finite non-translational parts in $\Sigma$ and $g$, which we write as:
\begin{equation}
g({\bf k},{\bf q},i\omega_{n}) = 
g({\bf k},i\omega_{n}) N_{c} \delta({\bf q}) +
\delta g({\bf k},{\bf q},i\omega_{n})\:,
\end{equation}
\begin{equation}
\Sigma({\bf k},{\bf q}) = 
\Sigma({\bf k}) N_{c} \delta({\bf q}) +
\phi({\bf k},{\bf q})\:.
\end{equation}
In order to study the boundary of the incommensurate phase, it
is sufficient to linearize Dyson's equation with respect to the 
non-translational parts, yielding

\begin{equation}
\delta g({\bf k},{\bf q},i\omega_{n}) =
g_{0}({\bf k},i\omega_{n})
\phi({\bf k},{\bf q})
g_{0}({\bf k-q},i\omega_{n})\:.
\label{lineariun}
\end{equation}
From Eq.(\ref{selg}), calculated at large $N$'s,
we obtain
the relation between $\phi$ and $\delta g$:
\begin{eqnarray}
\phi({\bf k},{\bf q}) & = &
\frac{1}{N_{c}} \sum_{{\bf p}} B({\bf k},{\bf q},{\bf p}) 
T\sum_{n} \delta g({\bf p},{\bf q},i\omega_{n})
e^{i\omega_{n} 0^{+}}=\nonumber\\
\mbox{} & = & 
\frac{1}{N_{c}} \sum_{{\bf p}} B({\bf k},{\bf q},{\bf p})
T\sum_{n}
g_{0}({\bf p},i\omega_{n})
g_{0}({\bf p-q},i\omega_{n})
\phi({\bf p},{\bf q})\:.
\label{selfcon1}
\end{eqnarray}

The kernel $B({\bf k},{\bf q},{\bf p})$
can be written as a sum of separable kernels:
\begin{eqnarray}
B({\bf k},{\bf q},{\bf p}) & = &
t({\bf k- q}) +t({\bf p}) +V({\bf q})-\frac{J({\bf q})}{2}
 -\frac{J({\bf k+p})}{2} =\nonumber\\
& =& \sum_{\alpha=1}^{6} F_{\alpha}({\bf k},{\bf q})
G_{\alpha}({\bf p},{\bf q})\:,
\end{eqnarray}
where
\begin{equation}
\vec{F}({\bf k},{\bf q})=\left[
t({\bf k- q}), 1, 
J \cos(k_{x}), J \sin(k_{x}),
J \cos(k_{y}), J \sin(k_{y})\right]\:,
\label{fff1}
\end{equation}
\begin{equation}
\vec{G}({\bf k},{\bf q})=\left[
1, t({\bf k}) +V({\bf q})-\frac{J({\bf q})}{2},
-\cos(k_{x}), -\sin(k_{x}),
-\cos(k_{y}), -\sin(k_{y})\right].
\label{ggg1}
\end{equation}
The general linearized solution of the non-translational 
self-energy can then be written as
\begin{equation}
\phi({\bf k},{\bf q}) =
\sum_{\alpha} f_{\alpha}({\bf q}) 
F_{\alpha}({\bf k},{\bf q})\:.
\label{ff1}
\end{equation}
Inserting Eq. (\ref{ff1}) in Eq. (\ref{selfcon1}), we obtain the
eigenvalue equation
\begin{equation}
\sum_{\beta}
\left[ \delta_{\alpha\beta} - a_{\alpha\beta}({\bf q})
\right] f_{\beta}({\bf q}) = 0\:,
\label{homo1}
\end{equation}
where the matrix elements $a_{\alpha\beta}({\bf q})$ in 
Eq. (\ref{homo1}) are defined by
\begin{equation}
a_{\alpha\beta}({\bf q}) = 
\frac{1}{N_{c}}\sum_{{\bf p}}
T \sum_{n} 
G_{\alpha}({\bf p},{\bf q})
F_{\beta}({\bf p},{\bf q})
g_{0}({\bf p},i\omega_{n})
g_{0}({\bf p-q},i\omega_{n})\:.
\end{equation}
The boundary of the incommensurate phase is determined
by the first onset of a non-trivial solution of the
homogeneous system Eq.(\ref{homo1}). 
>From a different point of view, such instabilities correspond to 
divergencies of the charge vertex\cite{zeyher2,zeyher4}.
The particular ${\bf q}_{c}$ 
where such divergencies occur determines the incommensurability vector.

The undoped case has been studied in detail \cite{affleck2}.
Due to the Fermi surface topology, all instabilities
occur for a commensurate wave vector, and the ground state
can be described as a $(\pi,\pi)$ flux phase
with flux $\pi$ per plaquette.
In our framework, the flux state is characterized 
by an order parameter Eq.(\ref{ff1}) which is imaginary and 
the antisymmetric combination of
its third and fourth component, i.e., 
\begin{equation}
\phi_{FL}({\bf k},{\bf Q}_{c}) \propto i[\cos(k_{x})-\cos(k_{y})]\:.
\label{para}
\end{equation}
Consequently, a $d$-wave gap is opened at half-filling of the band.
Another state of interest which, however, has a  higher energy at $\delta=0$,
is the socalled ``kite'' phase, which is twofold degenerate and described by
the order parameter 
$\phi_{KI}({\bf k},{\bf Q}_{c}) \propto i[\sin(k_{x}) \pm 
\sin(k_{y})]$.

It has been argued that for a finite doping both the flux
plaquette and the instability or modulation vector ${\bf q}_{c}$
are functions of the doping \cite{lederer,poilblanc,morse}.
In order to investigate this question in more detail, we have studied
instabilities of the normal state with respect
to all possible order parameters Eq.(\ref{ff1}), which can be constructed from
the
eigenvectors of the $6\times6$ matrix in Eq. (\ref{homo1}).
We found that the instability vector ${\bf q}_{c}$ which can be restricted 
to the irreducible Brillouin zone corresponding to 1/8 of the total
Brillouin zone always lies
in the $(0,\pi)$-$(\pi,\pi)$ direction, going
smoothly to $(\pi,\pi)$ for zero doping. 
\begin{figure}[t]
\centerline{\psfig{figure=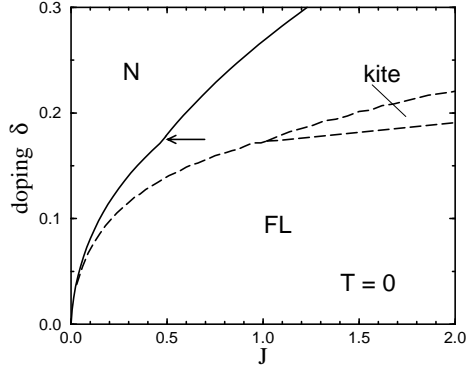,width=8cm}}
\caption{Solid line: Zero-temperature
boundary between normal state N and incommensurate flux state FL. The arrow
marks a transition between an incommensurate  flux at low and an 
incommensurate kite phase at high dopings. Dashed lines: Boundaries 
between normal state N and a commensurate flux or kite state.}
\label{figca1}
\end{figure}
Fig. 1 shows the calculated
phase diagram in the $\delta-J$ plane at zero temperature. From now on
we put $t=1$ so that all energies like $J$ are measured in units of
$t$. The dashed lines describe transitions to commensurate states
disregarding competing incommensurate states. For $J<1$ there is a
transition between the  normal state at large dopings to a commensurate
flux state at low dopings. 
For $J>1$ the normal state is with decreasing
doping first unstable with respect to the kite phase and the kite phase then
with respect to a flux state. Allowing also for incommensurate states the
solid line in Fig. 1 shows the boundary of the normal state with 
respect to an incommensurate flux state at $J<0.5$ and the kite phase
at $J>0.5$ where the exact transition point between the two incommensurate
states have been marked by an arrow. For $J>0.5$ there is probably another
solid line describing a transition between an incommensurate kite and
flux phase similar as in the commensurate case. Since we will confine
ourselves in the following to the parameter range 
$J<0.5$ relevant for high-$T_c$ superconductors we have not tried to
calculate this additional phase boundary.

Fig. 2a) characterizes the
components of the order parameter along the incommensurate boundary of 
Fig. 1. The labels 3,4,5,6 correspond to $\alpha=3,4,5,6$
in $F_\alpha$ defined in Eq.(\ref{fff1}). For dopings below
$\sim 0.175$ the order parameter is $\phi \sim \cos(k_x)-\cos(k_y)$
and thus has $d$-wave or, using the proper point group classification,
$\Gamma_3$ symmetry. Strictly speaking, $\phi$ has also a small
additional term $\propto  \sin(k_y)$ which may occur in incommensurate
but not in commensurate states. For dopings above $\delta \sim 0.175$ we have
$\phi \propto \sin(k_x)$ and thus an order parameter with $\Gamma_5$ symmetry. 
Writing the 
\begin{figure}[h]
\centerline{\psfig{figure=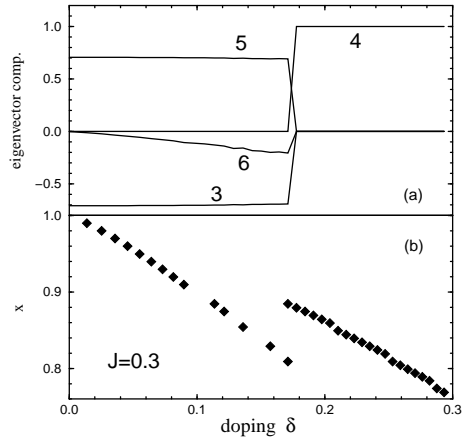,width=8cm,angle=270},clip=!}
\caption{(a) Evolution of the eigenvector components 
(3,4,5,6 correspond to $\cos(k_x)$, $\sin(k_x)$, $\cos(k_y)$, $\sin(k_y)$,
respectively), along the incommensurate boundary in Fig. 1;
(b) Corresponding dependence 
of the instability vector ${\bf q}_{c} = (1,x) \pi$.}
\label{figca2}
\end{figure}
instability
vector as ${\bf q}_c = (1,x)\pi$ the dependence of the value $x$
on the doping is shown in Fig. 2b). ${\bf q}_c$ approaches the
commensurate wave vector ${\bf Q}_c$ at zero doping. With increasing
doping it moves slowly away from ${\bf Q}_c$ and exhibits a jump
at $\delta \sim 0.175$ where the flux state is replaced by the kite
state. For the experimental value $J/t \sim 0.3$ only the flux
phase is possible and is the stable phase for dopings below $\delta \sim
0.13$.

\begin{figure}[t]
\centerline{\psfig{figure=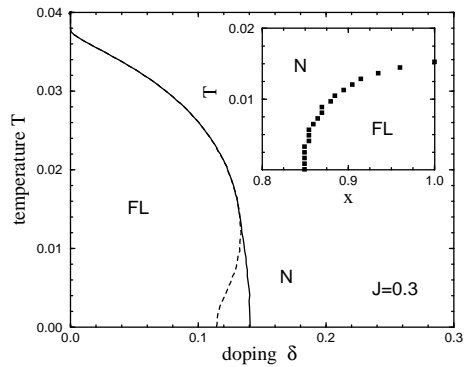,width=8cm,angle=270}}
\caption{Incommensurate (solid line) and commensurate (dashed line) 
phase boundaries in the $T-\delta$ plane. Insert: Evolution of
the instability vector ${\bf q}_{c} = (1,x) \pi$ as function of $T$.}
\label{figca3}
\end{figure}

Fig. 3 shows the phase diagram in the $T-\delta$ plane for $J=0.3$.
The solid line describes again the incommensurate, the broken line the
commensurate state. The inset of that Figure shows the temperature
dependence of the quantity $x$ of the instability vector ${\bf q}_c =(1,x) \pi$.
For temperatures larger than $\sim 0.014$ ${\bf q}_c$ coincides with
${\bf Q}_c$ so that the broken and solid lines become identical.
This behavior is caused by the thermal width of the Fermi function:
though the gap in the one-particle density opens not exactly at the
chemical potential the thermal smearing around the Fermi surface is large
enough to lower the free energy by taking advantage of the high density of 
states corresponding to the half-filled case. In the following we will assume
that the flux state is commensurate. This assumption simplifies considerably
calculations which also take superconducting states into account. On the
other hand Fig. 3 suggests that this is not an unreasonable approximation.

\section{Competition between superconductivity\newline
and flux phases}
\label{supersec}

In this section we consider instabilities of the normal and of the flux
state with respect to superconductivity. Superconducting instabilities
occur at most in $O(1/N)$ of our $1/N$ expansion. This means that it
is sufficient to calculate the Green's functions entering the linearized
gap equation for superconductivity in the leading order $O(1)$. A
detailed study of the normal to superconducting transition showed 
that the usual mean field term 
dominated all the other contributions to the kernel of the gap
equation. So we will keep just this term 
and neglect all the other terms which are also of of $O(1/N)$.
Moreover, it was shown that the leading superconducting instability
occurs in the $d$-wave-like $\Gamma_3$ channel so we will consider only
a $d$-wave order parameter. With respect to the order parameter matrix
$\phi$ we will keep the component $\propto \cos(k_x)-\cos(k_y)$
discussed in the last section. The usual charge-density wave order
parameter is according to Eq.(\ref{ff1}) connected to the columns 1 and 2
and the sum of 3 and 5
and will be included in the next section. Finally, the components
$\propto \sin(k_x)$ and $\propto \sin(k_y)$ are associated with the kite
phase and do not play a role for $J<0.5$ as shown in the previous
section. \par
In the presence of a commensurate flux phase and singlet superconductivity 
the operators of the following row vector
\begin{equation}
\psi({\bf k}) = (X^{0\sigma}({\bf k}),X^{{\bar \sigma}0}(-{\bf k}),
X^{0\sigma}({\bf k}-{\bf Q}_c),X^{{\bar \sigma}0}({\bf k}-{\bf Q}_c)),
\label{row}
\end{equation}
are coupled in the Green's functions. We thus define a $4\times4$ matrix
Green's function
\begin{equation}
\hat{G}(k) = -<TS\psi^\dagger(k)\psi(k)>/<S>,
\label{grow}
\end{equation}
where $\psi^\dagger$ is the hermitian conjugate of $\psi$, i.e.,
a column vector. $k$ in Eq.(\ref{grow}) denotes the four-component vector
$k=({\bf k},i\omega_n)$. We also note that in the absence of magnetic
ground states the Green's functions $\hat{G}$ are spin-independent
which allows to drop all spin indices. Going then over to normalized
Green's functions we  
write the $4\times4$ matrix $\hat{g}(k)$ as
\begin{equation}
\hat{g}(k) = \left(
\begin{array}{cccc}
 g_{11}(k) & g_{12}(k) 
 & g_{13}(k) & 0\\
 g_{21}(k) & 
 g_{22}(k) & 0 & 
 g_{24}(k)\\
 g_{13}(k-q_c) & 0 & 
 g_{11}(k-q_c) &
 g_{12}(k-q_c)\\
 0 & g_{24}(k-q_c) & 
 g_{21}(k-q_c) &
 g_{22}(k-q_c)
\end{array}
\right).
\label{gr4}
\end{equation}
Here we used the fact that the matrix elements 
$g_{14}, g_{23}, g_{32}, g_{41}$ of 
$\hat{g}$
must vanish and the explicite expression (\ref{row}) to connect
different matrix elements. 
We also used the four-component vector $q_c=({\bf Q}_c,0)$.
Dyson's equation becomes a $4\times4$ matrix equation

\begin{equation}
\left[ i\omega_{n} \hat{I} -\hat{\Sigma}(k)\right]
\cdot \hat{g}(k) = \hat{I}\:.
\label{dysonmatr}
\end{equation}
The self-energy matrix $\hat{\Sigma}$ has the general form

\begin{equation}
\hat{\Sigma}({\bf k}) = \left(
\begin{array}{cccc}
 \Sigma_{11}({\bf k}) & 
 \Delta_{12}({\bf k}) &
 i\phi_{13}({\bf k}) & 0\\
 \Delta_{21}({\bf k}) & 
 \Sigma_{22}({\bf k}) & 0 &
 i\phi_{24}({\bf k})\\
 i\phi_{13}({\bf k-Q}_c) & 0 & 
 \Sigma_{11}({\bf k-Q}_c) &
 \Delta_{12}({\bf k-Q}_c)\\
 0 & i\phi_{24}({\bf k-Q}_c) & 
 \Delta_{21}({\bf k-Q}_c) &
 \Sigma_{22}({\bf k-Q}_c)
\end{array}
\right).
\label{se4}
\end{equation}
Explicit expressions for the elements of $\hat{\Sigma}$ are obtained
from Eq.(\ref{selg}):

\begin{equation}
\Sigma_{ii}({\bf k}) = (-1)^{i+1}(1-n) 
\:\epsilon_{0}({\bf k}) 
-\alpha_{ii}({\bf k})
+\lambda_{ii} -\mu\:,
\label{sen}
\end{equation}
with
\begin{equation}
n = \frac{2}{N_{c}} \sum_{{\bf p}}  
T \sum_{n} 
g_{11}({\bf p},i\omega_{n}) e^{i\omega_{n}0^{+}}\,,
\label{nn}
\end{equation}
\begin{equation}
\lambda_{ii} = \frac{1}{N_{c}} \sum_{{\bf p}} 
\left[ t({\bf p}) - \frac{J({\bf q} = 0)}{2} + 
V({\bf q} \rightarrow 0) \right]
T \sum_{n} 
g_{ii}({\bf p},i\omega_{n}) e^{i\omega_{n}0^{+}}\,,
\end{equation}
\begin{equation}
\alpha_{ii}({\bf k}) = - \frac{1}{2 N_{c}} \sum_{{\bf p}} 
J({\bf k+p}) T \sum_{n} 
g_{ii}({\bf p},i\omega_{n}) e^{i\omega_{n}0^{+}}\,,
\label{endsen}
\end{equation}
and
\begin{equation} 
\Delta_{ij}({\bf k}) = 
- \frac{1}{2 N_{c}} \sum_{{\bf p}} 
[ J({\bf k+p}) - V_{d}({\bf k+p})]    T \sum_{n} 
g_{ij}({\bf p},i\omega_{n}) 
e^{i\omega_{n}0^{+}}\,,
\end{equation}
\begin{equation} 
i\phi_{ij}({\bf k}) = - \frac{1}{2 N_{c}} \sum_{{\bf p}} 
J({\bf k+p}) T \sum_{n} 
g_{ij}({\bf p},i\omega_{n}) e^{i\omega_{n}0^{+}}\,,
\label{sefl}
\end{equation}
where $V_{d}({\bf k})=2 V_{nn} [\cos(k_{x})+\cos(k_{y})]$ 
is the nearest neighbor part of the Coulomb interaction.
The wave vector ${\bf k}$ is either $\bf k$ or 
${\bf k}-{\bf Q}_c$, so that Eqs.(\ref{sen}) - (\ref{sefl}) determine all
matrix elements of the self-energy. In particular, all elements of 
$\hat{\Sigma}$
are independent of frequency in agreement with the notation in Eq.(\ref{se4}).
Eqs.(\ref{gr4}) and (\ref{se4}) determine all elements of $\hat{g}$ and 
$\hat{\Sigma}$.
It is easy to see that many of them are actually not independent.
Writing for the diagonal elements 
$\Sigma_{ii}({\bf k}) = \epsilon_{ii}({\bf k}) +\lambda_{ii}
-\mu$ we have $\lambda = \lambda_{ii}$ and
\begin{equation}
\epsilon({\bf k}) =
\epsilon_{11}({\bf k}) =
- \epsilon_{22}({\bf k}) =
- \epsilon_{11}({\bf k-Q}_{c}) =
\epsilon_{22}({\bf k-Q}_{c})\:.
\end{equation}
Similarly, one obtains for the elements of the order parameters

\begin{equation}
\Delta({\bf k}) =
\Delta_{12}({\bf k}) =
\Delta_{21}({\bf k}) =
- \Delta_{12}({\bf k-Q}_{c}) =
- \Delta_{21}({\bf k-Q}_{c})\:,
\label{rede}
\end{equation}
\begin{equation}
\phi({\bf k}) =
\phi_{13}({\bf k}) =
\phi_{24}({\bf k}) =
- \phi_{13}({\bf k-Q}_{c}) =
- \phi_{24}({\bf k-Q}_{c})\:.
\end{equation}
In Eq.(\ref{rede}) we have used the fact 
that $\Delta$ can be chosen to be real.

$\hat{\Sigma}({\bf k})$ can easily be diagonalized and one obtains
four branches for the excitation spectrum:

\begin{equation}
i\omega_{n} =
\pm E_{\pm}({\bf k}) = \pm \sqrt{[\xi({\bf k}) 
\pm (\mu - \lambda)]^{2}+\Delta({\bf k})^{2}}\:,
\end{equation}
with
\begin{equation}
\xi({\bf k}) = \sqrt{\varepsilon({\bf k})^{2}+
\phi({\bf k})^{2}}\:.
\end{equation}

Eq.(\ref{dysonmatr}) can therefore easily be inverted
and the sum over frequencies in Eqs.(\ref{nn})-(\ref{sefl})
carried out. We obtain the following set of self-consistent
equations:

\begin{eqnarray}
n & = & 1 +
\frac{2}{N_{c}} \sum_{{\bf p}}\mbox{}'
\left\{ 
\frac{\xi({\bf p})-\lambda+\mu}{2E_{+}({\bf p})}
\tanh\left[\frac{E_{+}({\bf p})}{2 T}\right]\right.
\nonumber\\
 & & \left. -
\frac{\xi({\bf p})+\lambda-\mu}{2E_{-}({\bf p})}
\tanh\left[\frac{E_{-}({\bf p})}{2 T}\right]\right\},
\label{n1}
\end{eqnarray}
\begin{equation}
\lambda = \frac{1}{N_{c}} \sum_{{\bf p}}\mbox{}'
t({\bf p})\, \eta({\bf p}) -\frac{1}{2}\left[\frac{J({\bf 0})}{2}-
V({\bf 0})\right] n   \:,
\label{lambda1}
\end{equation}
\begin{equation}
\alpha({\bf k}) = - \frac{1}{2 N_{c}} \sum_{{\bf p}}\mbox{}'
J({\bf k+p})\, \eta({\bf p})\:,
\label{alpha1}
\end{equation}
\begin{equation}
\Delta({\bf k}) = \frac{1}{2 N_{c}} \sum_{{\bf p}}\mbox{}'
[ J({\bf k+p}) - V_{d}({\bf k+p})]\, \eta_{\Delta}({\bf p})\:,
\label{delta1}
\end{equation}
\begin{equation}
\phi({\bf k}) =  \frac{1}{2 N_{c}} \sum_{{\bf p}}\mbox{}'
J({\bf k+p})\, \eta_{\phi}({\bf p})\:,
\label{phi1}
\end{equation}
where
\begin{equation}
\eta({\bf k}) =
- \frac{\varepsilon({\bf k})}{\xi({\bf k})}
\left\{
\frac{\xi({\bf k})-\lambda+\mu}{2E_{+}({\bf k})}
\tanh\left[\frac{E_{+}({\bf k})}{2 T}\right]+
\frac{\xi({\bf k})+\lambda-\mu}{2E_{-}({\bf k})}
\tanh\left[\frac{E_{-}({\bf k})}{2 T}\right]\right\},
\label{etan}
\end{equation}
\begin{equation}
\eta_{\Delta}({\bf k}) =
\left\{
\frac{\Delta({\bf k})}{2E_{+}({\bf k})}
\tanh\left[\frac{E_{+}({\bf k})}{2 T}\right]+
\frac{\Delta({\bf k})}{2E_{-}({\bf k})}
\tanh\left[\frac{E_{-}({\bf k})}{2 T}\right]\right\},
\end{equation}
\begin{equation}
\eta_{\phi}({\bf k}) =
\frac{\phi({\bf k})}{\xi({\bf k})}
\left\{
\frac{\xi({\bf k})-\lambda+\mu}{2E_{+}({\bf k})}
\tanh\left[\frac{E_{+}({\bf k})}{2 T}\right]+
\frac{\xi({\bf k})+\lambda-\mu}{2E_{-}({\bf k})}
\tanh\left[\frac{E_{-}({\bf k})}{2 T}\right]\right\},
\label{etaphi}
\end{equation}
and the prime on the summation indicates that the sum is 
restricted to the reduced Brillouin zone.

Eqs. (\ref{n1})-(\ref{phi1}), together with Eqs.
(\ref{etan})-(\ref{etaphi}), determine in a 
self-consistent way
all the properties of the system.
Moreover, it is possible to construct the thermodynamical
potential $\Omega(\mu,T)$  as that function which satisfies the extremal 
conditions
\begin{equation}
\frac{\delta \Omega}{\delta \alpha({\bf k})} = 0\,,
\hspace{15 pt}
\frac{\delta \Omega}{\delta \Delta({\bf k})} = 0\,,
\hspace{15 pt}
\frac{\delta \Omega}{\delta \phi({\bf k})} = 0\,,
\end{equation}
together with the conditions 
$\partial \Omega / \partial \lambda = 0$,
$- \partial \Omega / \partial \mu = n$. Performing a Legendre transformation
on $\Omega$ one obtains the following expression for the free energy
$F(N,T)$:
\begin{eqnarray}
F(N,T) & = & (\mu - \lambda) (n - 1)\nonumber\\
\mbox{} & \mbox{} & 
-2 T \sum_{{\bf k}}\mbox{}' \left\{
\ln\left[ 2 \cosh\left(\frac{E_{+}({\bf k})}{2 T}\right)\right]
+\ln\left[ 2 \cosh\left(\frac{E_{-}({\bf k})}{2 T}\right)\right]
\right\}\nonumber\\
\mbox{} & \mbox{} & 
+ \frac{1}{2N_{c}^{2}} \sum_{{\bf k},{\bf p} }\mbox{}'
J({\bf k+p}) \eta({\bf k}) \eta({\bf p})
+ \frac{1}{2 N_{c}^{2}} \sum_{{\bf k},{\bf p} }\mbox{}'
J({\bf k+p}) \eta_{\phi}({\bf k}) \eta_{\phi}({\bf p})
\nonumber\\
\mbox{} & \mbox{} & 
+ \frac{1}{2N_{c}^{2}} \sum_{{\bf k},{\bf p} }\mbox{}'
[J({\bf k+p})-V_{d}({\bf k+p})] 
\eta_{\Delta}({\bf k}) \eta_{\Delta}({\bf p})\:.
\end{eqnarray}

After having derived the system of equations for the order parameters
we are going to analyze the phase diagram of the $t-J$ model within 
our approach. We choose the generally accepted value $J=0.3$ 
using always $t$ as the energy unit. An estimate
of the Coulomb repulsion between nearest neighbor sites, $V_{nn}$,
may be more controversial. It seems reasonable to assume that
$V_{nn}$ is of the same order as $J$, so we have chosen the value
$V_{nn} = 0.5 J$. This particular choice means that the contribution
from density-density interactions in the $J$ and $V_{nn}$ terms
cancel each other in the $d$-wave superconducting channel,
so that superconductivity is driven only
by spin-exchange\cite{greco}.
The role played by $V_{nn}$ and the Coulomb interaction in the phase diagram
will be discussed in more detail in the next section.

\begin{figure}[t]
\centerline{\psfig{figure=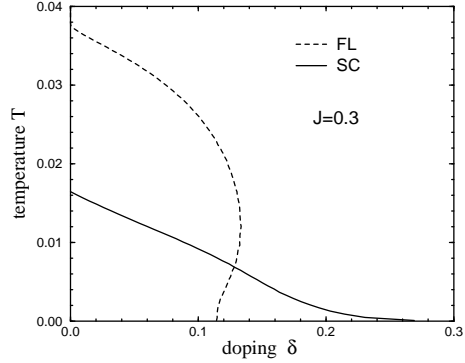,width=8cm,angle=270}}
\caption{Phase diagram of the $t$-$J$ model for
decoupled $d$-wave superconducting (SC) and commensurate flux (FL) order 
parameters.}
\label{figca4}
\end{figure}

Fig. \ref{figca4} shows the instability line of the normal state
with respect to $d$-wave superconductivity (solid line) and commensurate
flux phase (dashed line) in the $T-\delta$ plane assuming that the
two phases are uncoupled. $T_c$ of the uncoupled superconducting
phase increases with decreasing doping $\delta$. This behavior is
caused mainly by an increase of the density of states with decreasing
$\delta$ due to the narrowing of the band and the decreasing
distance of the Fermi energy from the Van Hove singularity at the middle
of the band. The transition temperature $T_{FL}$ to the flux phase 
becomes non-zero at around $\delta \sim 0.13$ and assumes very
rapidly large value towards lower dopings. Though this instability is
caused by nesting properties of the quasi two-dimensional Fermi surface
the dashed line in Fig. \ref{figca4} indicates that it is much stronger than
the instability towards superconductivity. Both phases have order parameters
of $d$-wave symmetry, i.e., nodes along the [1,1]- and maximum absolute
values along the [1,0]- and [0,1]-directions in ${\bf k}$-space. Both
phases thus try to reconstruct the Fermi surface mainly around the $X-$ and
$Y-$ points creating there a gap in the single-particle excitation
spectrum. In the case of superconductivity the gap opens always right at
the Fermi energy and moves with doping. In contrast to that the gap of
the flux phase is fixed at the middle of the gap due to the assumed
commensurability. Nevertheless, there will be a large competition
between the two phases for not too large dopings. Fig. \ref{figca4}
suggests that the flux phase is able to reconstruct also electronic states
further away from the Fermi surface compared to the superconducting 
phase which affects mainly electronic states close to the Fermi energy.

Fig. 5 shows the phase diagram of the $t-J$ model if the
interaction between the flux and superconducting phases is taken into
account. For $\delta > \sim 0.13$ only the superconducting phase is
stable at low temperatures. At around $\delta = 0.13$ the flux phase
order parameter becomes non-zero in the superconducting phase 
and a coexistence region exists of superconductivity and flux phase.
Because the two order parameters have the same symmetry and aim to
reconstruct the same parts of the Fermi surface the stronger of the
two phases tries to suppress the weaker one. Since $T_{FL} > T_c$
for $\delta < \sim 0.13$ the superconducting phase is rapidly suppressed
for decreasing dopings. As a result the solid line in Fig. \ref{figca4}
which increases monotonously with decreasing doping tends
rapidly to zero below $\delta \sim 0.13$ 
due to the interaction with the flux phase. The maximum value for $T_c$
coincides rather accurately with the onset of the flux phase
at $T=0$. Fig. \ref{figca6} shows the flux per plaquette $\Phi$ in the
interacting case as a function of doping for three different temperatures.
The flux always assumes its maximum value of $\pi$ at zero
doping. With increasing doping or with increasing temperature $\Phi$
decays rather fast. In the coexistence region of superconductivity and
flux phase $\Phi$ is small but non-zero. 

\begin{figure}[t]
\centerline{\psfig{figure=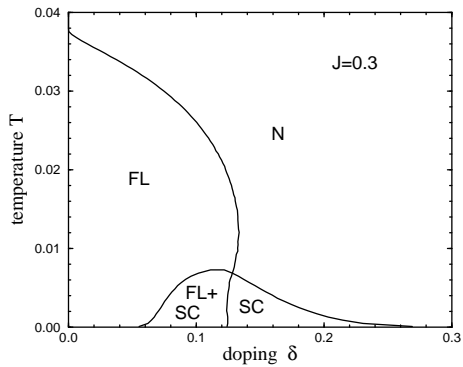,width=8cm,angle=270}}
\caption{Phase diagram of the $t$-$J$ model taking into
account the competition between superconductivity
and commensurate flux phase.}
\label{Tinterplay}
\end{figure}

Both Figs. \ref{figca3} and \ref{figca6} show that at low temperatures
$T < \sim 0.01$ an incommensurate flux state is more stable than 
the commensurate one. In an incommensurate flux state the one-particle
gap opens like in the case of superconductivity right at the Fermi
energy. From this one might expect an even larger competition between
flux and superconductivity phases and a more rapid quenching of $T_c$
at low dopings than in the case of a commensurate flux phase. 
We thus arrive in a natural way at a scenario with a
quantum critical point. 
Omitting superconductivity 
the metallic state at large dopings passes at zero temperature with 
decreasing doping through a 
critical point into a non-metallic flux state characterized by
an order parameter with $d$-wave symmetry and an incommensurate modulation
vector. At finite temperatures the long-range order of the flux state is
destroyed but there are still regions
of small and large fluctuations in the flux order parameter. Allowing also
for superconductivity the surroundings of the quantum critical point  
becomes superconducting with a $T_c$ which has a maximum at the critical
point and decays rapidly at larger or smaller dopings. One attractive
feature of our phase diagram is the near coincidence of critical point and
maximal $T_c$
\begin{figure}[h]
\centerline{\psfig{figure=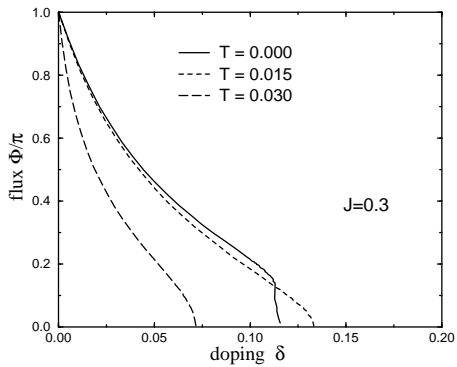,width=8cm,angle=270}}
\caption{The flux $\Phi$ per plaquette as a function of
doping for different temperatures.}
\label{figca6}
\end{figure}
which is a necessary consequence of the competition of
flux and superconducting phases. This coincidence seems to 
be a generic
feature of high-$T_c$ oxides and does not rely in our approach on parameter 
choices or fine-tuning.

\section{Coulomb interaction and incommensurate\newline
char\-ge-density-wave}
\label{pssec}

We have seen in section II that macroscopic phase separation is impossible
in the presence of long-range Coulomb forces. However,
it is known that in such a  situation the charge instability
at ${\bf q}=0$ corresponding to phase separation is shifted to a
finite ${\bf q}$, leading to an incommensurate charge-density-wave 
state\cite{castellani2}. Commensurate (CDW)\cite{markiewicz} 
or incommensurate charge-density-wave states (ICDW)\cite{castellani1,becca}
have been proposed to account for the metal-insulator transition at
$T=0$, the quantum critical point, and the pseudo-gap features at finite
temperatures in the underdoped regime of high-$T_c$ oxides. 
Though we have proposed a different candidate for the insulating state 
in the previous section the $t-J$ model may exhibit charge separation
and ICDW's in addition to the discussed flux phase. Thus the phase
digram in Fig. 5 may have to be modified.
We therefore study in the following
possible CDW instabilities including in the Hamiltonian also the
long-range Coulomb interaction $V({\bf q})$ defined in Eq.(\ref{v(q)}).
For this we consider a general ground state characterized in general
by nonvanishing order parameters for $d$-wave superconductivity and
a commensurate $d$-wave flux state. The corresponding Green's function and
self-energies form $4\times4$ matrices as has been discussed in section IV. Atop
on this ground state we allow for a small ICDW and check whether its
amplitude can be nonzero. The procedure is similar to that used in
section III for an incommensurate flux state. 

We write the Green's function and the self-energy as

\begin{equation}
\hat{g}({\bf k},{\bf q},i\omega_{n}) = 
\hat{g}({\bf k},i\omega_{n}) N_{c} \delta({\bf q}) +
\delta\hat{g}({\bf k},{\bf q},i\omega_{n})\:,
\label{glin}
\end{equation}
\begin{equation}
\hat{\Sigma}({\bf k},{\bf q}) = 
\hat{\Sigma}({\bf k}) N_{c} \delta({\bf q}) +
\hat{\phi}_{CDW}({\bf k},{\bf q})\:.
\label{slin}
\end{equation}
The first term on the right-hand sides of 
Eqs.(\ref{glin}) and (\ref{slin}) 
describe the state with superconductivity and commensurate flux phase,
both with $d$-wave symmetry. $\hat{\phi}_{CDW}$ and $\delta \hat{g}$ 
are small non-translational additions to $\hat{\Sigma}$ and $\hat{g}$
due to the ICDW. It is sufficient to linearize Dyson's equation
with respect to the non-translational parts yielding

\begin{equation}
\delta\hat{g}({\bf k},{\bf q},i\omega_{n}) =
\hat{g}({\bf k},i\omega_{n})\cdot
\hat{\phi}_{CDW}({\bf k},{\bf q})
\cdot\hat{g}({\bf k-q},i\omega_{n})\:.
\label{lineari}
\end{equation}
Since we are dealing with a CDW instability 
$\hat{\phi}_{CDW}({\bf k},{\bf q})$ is diagonal 
in the $4\times4$ space and can be written as a linear combination of the 
two $4\times4$ matrices:
\begin{equation}
\hat{\tau_3} = 
\left(
\begin{array}{cc}
 \hat{\sigma}_{3}  & \hat{0}\\
 \hat{0} & - \hat{\sigma}_{3}
\end{array}
\right),
\end{equation}
\begin{equation}
\hat{\tau_0} = 
\left(
\begin{array}{cc}
 \hat{\sigma}_{3}  &  \hat{0}\\
 \hat{0} & \hat{\sigma}_{3}
\end{array}
\right),
\end{equation}
where $\hat{\sigma_3}$ is the usual third Pauli matrix. Explicit calculations 
of the diagonal elements of the self-energy using Eq.(\ref{selg}) 
yield, similar as in Eq.(\ref{selfcon1}), the result

\[
\hat{\phi}_{CDW}({\bf k},{\bf q}) 
= 
\hat{\tau_3}  t({\bf k-q})\frac{1}{N_{c}}
\sum_{{\bf p}}\mbox{}'  
T \sum_{n} 
{\rm Tr} \left[\hat{\tau_0}  \cdot 
\delta\hat{g}({\bf p},{\bf q},i\omega_{n}) \right]
\]
\[
+  \hat{\tau_0}  \frac{1}{N_{c}}\sum_{{\bf p}}\mbox{}'
t({\bf p})  
T \sum_{n} 
{\rm Tr} \left[ \hat{\tau_3} \cdot 
\delta \hat{g}({\bf p},{\bf q},i\omega_{n}) \right]
\]

\[
+  \hat{\tau_0}  
( -\frac{J({\bf q})}{2}+V({\bf q})) \frac{1}{N_{c}} 
\sum_{{\bf p}}\mbox{}'
T \sum_{n} 
{\rm Tr} \left[ \hat{\tau_3} \cdot 
\delta \hat{g}({\bf p},{\bf q},i\omega_{n}) \right]
\]
\begin{equation}
-  \hat{\tau_3}
\frac{1}{2 N_{c}}\sum_{{\bf p}}\mbox{}'
J({\bf k+p})
T \sum_{n} 
{\rm Tr} \left[ \hat{\tau_3} \cdot 
\delta\hat{g}({\bf p},{\bf q},i\omega_{n}) \right]
\:.
\label{selffluxicdw}
\end{equation}
Eq.(\ref{selffluxicdw}) can be written in the more compact form

\begin{equation}
\hat{\phi}_{CDW}({\bf k},{\bf q}) = 
\frac{1}{N_{c}}\sum_{{\bf p}}\mbox{}'
T \sum_{n} 
\sum_{\alpha} \hat{F}_{\alpha}({\bf k},{\bf q})
{\rm Tr} \left[
\hat{G}_{\alpha}({\bf p},{\bf q})\cdot
\delta \hat{g}({{\bf p},{\bf q},},i\omega_{n}) \right],
\label{selfcon}
\end{equation}
\begin{figure}[t]
\centerline{\psfig{figure=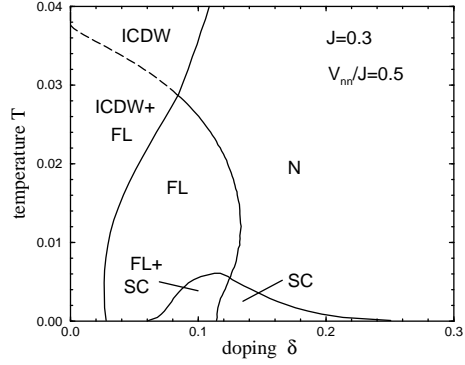,width=8cm,angle=270}}
\caption{Phase diagram of the $t$-$J$ model, considering also an
in\-com\-men\-su\-ra\-te charge-density-wave (ICDW). The dashed line
between ICDW and ICDW+FL is purely indicative.}
\label{figca7}
\end{figure}
with the abbreviations

\begin{equation}
\hat{F}({\bf k},{\bf q}) =
\left[ t({\bf k-q}) \hat{\tau_3},
\hat{\tau_0},
\frac{J({\bf k})}{2} \hat{\tau_3} \right],
\end{equation}
\begin{equation}
\hat{G}({\bf k},{\bf q}) =
\left[ \hat{\tau_0},
t({\bf k}) \hat{\tau_3} + 
\left[ V({\bf q})-\frac{J({\bf q})}{2}\right]
\hat{\tau_0},
- \frac{t({\bf k})}{4}\hat{\tau_3} \right].
\end{equation}
Similar as in section \ref{incomsec},
the solution of the ICDW order parameter 
can be written as
\begin{equation}
\hat{\phi}_{CDW}({\bf k},{\bf q}) =
\sum_{\alpha} f_{\alpha}({\bf q}) 
\hat{F}_{\alpha}({\bf k},{\bf q})\:.
\label{ff}
\end{equation}
Using Eq.(\ref{lineari}) the expansion coefficients $f_\alpha$ satisfy the 
following system of 3 equations

\begin{equation}
\sum_{\beta}
\left[ \delta_{\alpha\beta} - a_{\alpha\beta}({\bf q})
\right] f_{\beta}({\bf q}) = 0\:,
\label{homo}
\end{equation}
where the matrix elements $a_{\alpha\beta}({\bf q})$ 
are defined by
\begin{equation}
a_{\alpha\beta}({\bf q}) = 
\frac{1}{N_{c}}\sum_{{\bf p}}\mbox{}'
T \sum_{n} 
{\rm Tr} \left[
\hat{G}_{\alpha}({\bf p},{\bf q})\cdot
\hat{g}_{0}({\bf p},i\omega_{n})\cdot 
\hat{F}_{\beta}({\bf p},{\bf q})\cdot
\hat{g}_{0}({\bf p-q},i\omega_{n})
\right].
\end{equation}
As usual the phase boundary is determined
by the onset of the first non-trivial solution of the
homogeneous system Eq.(\ref{homo}). The particular ${\bf q}_{c}$ 
where this occurs determines the incommensurability vector
of the ICDW.

Taking also ICDW's into account the calculated phase diagram is shown
in Fig. 7 using $J=0.3$ and $V_{nn}/J = 0.5$. We also used the
long-range Coulomb potential Eq.(\ref{v(q)}) in the calculation. We characterize
its strength by its value $V_{nn}$ between nearest-neighbor sites. 
Fig. 7 should be compared with Fig. 5 where the same parameter values
have been used but the Coulomb interaction was confined to nearest neighbors.
The curves for $T_c$ and $T_{FL}$ are very similar in both cases. 
The region of the
flux phase is now, however, split into two regions. In one region
at larger dopings only the flux phase is stable; in the other one
at lower dopings the flux and the incommensurate charge density phase 
coexist with each other. At high temperatures the pure ICDW is stable 
at lower dopings.
The Figure indicates that the ICDW has no important influence on 
the $T_c$ curve and, in particular, to its maximum value at optimal doping. 
The latter is still determined solely by the instability towards the flux
phase at around $\delta = 0.13$.

\begin{figure}[t]
\centerline{\psfig{figure=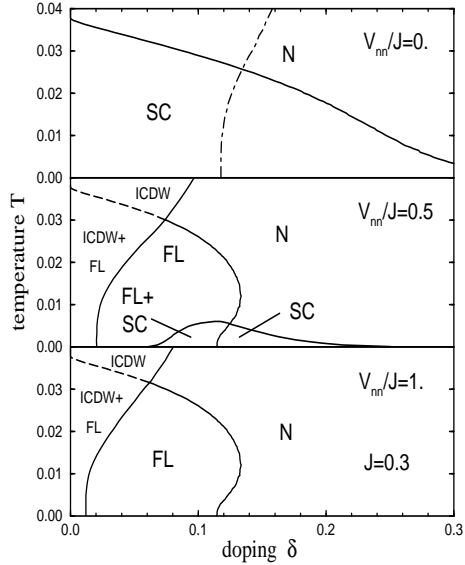,height=9cm,width=8cm,angle=270}}
\caption{Evolution of the phase diagram for $J=0.3$ and different Coulomb 
interaction
strengths, characterized by the nearest-neighbor constant $V_{nn}$.
For $V_{nn}=0$ the ICDW instability occurs at
${\bf q}_{c}=0$ corresponding to the divergence of the isothermal
compressibility and the onset of phase separation (dot-dashed line).}
\label{figca8}
\end{figure}
Finally we discuss the dependence of the phase diagram on the Coulomb 
repulsion strength $V_C$, defined in Eq. (\ref{v(q)}). Similar as in Fig. 7
we use instead of $V_C$ the value $V_{nn}$, i.e., the Coulomb potential 
between nearest neighbor sites, to characterize the strength of the Coulomb 
potential. Changing $V_{nn}$ we can distinguish between two extreme limits.
In the case of negligible Coulomb repulsion, i.e. $V_{nn}=0$, the
attractive charge-charge term of the $t-J$ model becomes important.
As a result the superconducting region becomes large and wipes out the flux
phase as shown in the upper panel of Fig. 8. Lowering the temperature
from high values one crosses the solid line and enters the superconducting
region. The phase boundary between normal and flux phase in Fig. 5
lies now within the superconducting region where according to the calculation
this boundary no longer exists. This can be understood from the fact
that the superconductivity order parameter has already reconstructed the
Fermi surface, especially near the points $X$ and $Y$, so that an
additional order parameter with $d$-wave symmetry cannot lower further the free
energy. At zero temperature the ground state is always superconducting 
in agreement with the arguments of Ref. \onlinecite{zhang}. Exactly at zero 
doping the
superconducting and the flux phases become equivalent, again in agreement
with previous arguments. The isothermal  compressibility diverges
along the dot-dashed line in the upper panel of Fig. 8. This means
that on the left side of this line the normal and superconducting states
are unstable as homogeneous phases, and the system 
separates into two phases, one with zero and the other with a high hole
density. \par

Increasing $V_{nn}$ from zero to finite values the superconducting
state is more and more suppressed whereas the boundary to the flux phase 
is unaffected.
This is shown in the middle panel of Fig. 8 for $V_{nn}/J=0.5$ which is
the same diagram as Fig. 7. Moreover the superconducting region splits up into
a pure superconducting part at larger dopings and a region at lower
dopings where the superconducting and the flux order parameters coexist.
When $V_{nn}$ exceeds the value $J$ the total effective hole-hole
interaction becomes repulsive and superconductivity is totally suppressed.
At the same time the instability of the flux phase with respect to
an additional ICDW moves towards smaller dopings, i.e., the region
of the pure flux phase increases also on the cost of the coexistence
region of flux and ICDW phases. This is illustrated in the lower panel in 
Fig. 8. Fig. 8 demonstrates, in particular, two things: with 
increasing $V_{nn}$ the CDW instabilities move monotonically to lower 
dopings. The position of these instabilities is in general
far away from optimal doping and thus 
does not influence much the region where superconductivity is the stable 
phase. For $V_{nn} > 0$ optimal doping is more or less determined   
by the onset of the flux phase at $T=0$ and thus tied to this instability.
\par
We are now in a position to make a comparison of our results with
those of other treatments. Ref. \onlinecite{sandalov} also enforces 
the constraint by
$X$-operators but does not find any instability of the normal state with
respect to a flux phase. Considering the case $N=2$ from the outset
Ref. \onlinecite{sandalov} uses a mean-field like decoupling procedure which 
violates 
Luttinger's theorem. For instance, the Fermi surface for $\delta = 1/3$
corresponds to half-filling in theories where Luttinger's theorem
is fulfilled. The absence of a flux phase instability of the normal state
as well as a finite $T_c$ for superconductivity even at $J=0$ may be 
artefacts due this short-coming. Ref. \onlinecite{onufrieva} calculates the 
expectation values
of bosonic Hubbard operators and the Green's functions with different 
perturbation expansions finding also solutions for the $n(\mu)$ relation
which satisfy Luttinger's theorem. The resulting $T_c(\delta)$ curves
are similar to ours but again flux phases and their competition with
superconductivity are missing. A basic inconsistency problem 
inherent in simple decoupling schemes with $X$-operators can also be
inferred from a comparison of 
Ref. \onlinecite{sandalov} with Ref. \onlinecite{onufrieva}:
Calculating expectation
values of bosonic Hubbard operators from Green's functions using the 
projection properties of Hubbard operators or from thermodynamic relations
yields different $n(\mu)$ relations. In contrast to that the
$1/N$ expansion yields a unique $n(\mu)$ relation and also satisfies
Luttinger's theorem. On the other hand we cannot say much about 
the convergence of the $1/N$ expansion in general. However, in the case
of the density fluctuation spectrum it has been shown \cite{zeyher4} that the 
leading order in $1/N$ can already account for most features found
in the spectra calculated by exact diagonalizations for $N=2$ for small 
systems and that
the remaining discrepancies nearly vanish if next-to-leading contributions
are also taken into account \cite{khalliulin}. \par
Finally we compare our results with treatments where the constraint is
enforced using slave particles. One general result of these approaches  
is \cite{zhang,wang} that the staggered flux phase is always unstable
at $T=0$ against $d$-wave superconductivity. This agrees with our
findings, see the upper panel of Fig. 8. Fig. 4 of Ref. \onlinecite{ubbens}
and Fig. 1 of Ref. \onlinecite{wen}, if scaled to our value $J/t=0.3$,
show at finite temperatures no or only a very small region at very small 
dopings where the flux phase is stable. In our case the flux phase is wiped
out either by superconductivity or by phase separation as shown in the upper
panel of Fig. 8. Taking also Coulomb interactions into account to
prevent macroscopic phase separation we find that the flux phase becomes 
stable above the superconducting phase yielding a maximal $T_c$
of $\delta \sim 0.12$. This value is much larger than the 
value $\delta \sim 0.03$ obtained in Ref. \onlinecite{wen} and also closer
to the experimental one of $\delta \sim 0.15$. The decrease of $T_c$
with decreasing $\delta$ in the underdoped region is determined in
Refs. \onlinecite{ubbens,wen} by the condensation temperature of the slave
boson particles. Such a Bose condensation does not exist in our approach.
Instead the decrease of $T_c$ in the underdoped region is caused in the
present approach by the competition of the two $d$-wave order parameters 
describing the flux and the superconductivity phase. We also note that
experimental data have been interpreted in a phenomenological way
as a competition of the superconducting and an unknown phase \cite{cooper}
and the resulting phase diagram is very similar to those of Figs. 5 and 8.
The leading order of the $1/N$ expansion
is certainly insufficient for a proper description of the undoped case
$\delta=0$. It is now generally accepted that the obtained resonance-valence
bond instability is in this case somewhat weaker than the instability
towards long-range antiferromagnetism. The latter instability, however,
can only be obtained by taking higher order contributions of the $1/N$ 
expansion into account.

\section{Conclusions}

In this paper  we have derived the phase diagram of a generalized
$t-J$ model taking superconducting, flux, and charge density wave states
into account. The investigation was based on the leading expressions
of a $1/N$ expansion enforcing the constraints by means of $X$-operators.
We found a strong competition between $d$-wave superconducting and 
$d$-wave flux states. As a result the transition temperate $T_c$
for superconductivity showed a maximum near a doping value $\delta
= \delta_c \sim 0.13$ for $J/t = 0.3$. This value is determined 
essentially by the onset of an (incommensurate) flux phase
at $\delta_c$. To simplify the calculations we assumed the flux phase
to be commensurate. We showed that this is correct for $T/t > 0.01$.
Below this temperature the flux phase is incommensurate which, if taken
into account, would presumably not change substantially
our conclusions. We also studied the influence
of long-range Coulomb forces on the phase diagram. As a result
incommensurate charge density waves become stable or coexist with the flux
phase at lower dopings well separated from the superconducting region
influencing the latter at most in a marginal way.  \par
Our results can be interpreted in terms of a quantum critical point scenario.
Disregarding superconductivity the metallic state at large dopings
$\delta > \delta_c$ passes to a non-metallic, incommensurate flux
state with $d$-wave symmetry for $\delta < \delta_c$. Allowing also
for superconductivity $T_c$ increases from the over- and underdoped 
sides and shows a maximum around $\delta_c$. On the underdoped side
the superconducting phase coexist with the flux phase up to $T_c$
where a pure flux state becomes the most stable state up to the normal
state at high temperatures. Two different
proposals for the non-metallic state in the quantum critical point
scenario have been made, namely an antiferromagnetic\cite{pines} 
and an incommensurate
charge density wave state\cite{castellani1,becca}. 
In comparison with these states we would like 
to point out three attractive
features of our proposal for the non-metallic state. The instability
towards an incommensurate flux phase is a generic feature of the
$t-J$ model and is also present if second-nearest neighbor hopping or
Coulomb forces are additionally taken into account. For $J/t<0.5$
the flux phase has $d$-wave symmetry, i.e., the same symmetry as the
most stable superconducting state. Since the flux phase instability
is much stronger than the superconducting one $T_c$ is heavily suppressed
by the flux phase in the underdoped regime. As a necessary consequence 
the maximum value for $T_c$ lies near the onset of the flux phase
at $\delta_c$.

\end{document}